# Self-Phasematched Nonlinear Optics in Integrated Semiconductor Microcavities

**Alex Hayat and Meir Orenstein**

Department of Electrical Engineering, Technion, Haifa 32000, Israel

meiro@ee.technion.ac.il

ahayat@tx.technion.ac.il

A novel concept of self-phasematched optical frequency conversion in dispersive dielectric microcavities is studied theoretically and experimentally. We develop a time-dependent model, incorporating the dispersion into the structure of the spatial cavity modes and translating the phasematching requirement into the optimization of a nonlinear cavity mode overlap. We design and fabricate integrated double-resonance semiconductor microcavities for self-phasematched second harmonic generation. The measured efficiency exhibits a significant maximum near the cavity resonance due to the intra-cavity enhancement of the input power and the dispersion-induced wavelength detuning effect on the mode overlap, in good agreement with our theoretical predictions.

Miniature photonic devices for second-order nonlinear optics including frequency converters [10,16], electro-optic modulators and parametric oscillators are of a great importance for integrated photonics and optical communications [1]. Furthermore, compact sources of single photons and entangled photon pairs [2], essential for quantum communications and quantum information processing, can be realized using spontaneous parametric down-converters [12, 15].

In most of the second-order nonlinear optics experiments, material dispersion is compensated by birefringent crystal phase-matching [2] or quasi phase-matching (QPM) in periodically poled ferroelectric crystals [3], where the nonlinear susceptibility $\chi^{(2)}$ is spatially modulated with a period shorter than the nonlinear process coherence length - Lc. Semiconductors offer a promising alternative due to their high nonlinear susceptibilities and compatibility with the existing photonic-circuit technology. The main limitation of semiconductor materials, however, is their optical isotropy, which inhibits natural birefringent phase matching, whereas implementation of QPM techniques in semiconductors [4] appears to be very difficult. A number of technologically promising methods were developed for semiconductor waveguide phasematching including form birefringence [10-12] which can be implemented using specific materials, and modal phasematching [13], where the efficiency is usually limited by overlap of the transverse waveguide modes.

Nonlinear wave mixing in dielectric cavities was recently proposed for materials which cannot be phase-matched by conventional means. A properly designed cavity with length shorter than Lc was shown to provide significant nonlinear process efficiency in integrated semiconductor microcavities by phasematching the traveling-waves with dispersive dielectric mirrors [5] or using ring resonators [14]. Moreover, high quality factor cavities can enhance the efficiency by raising the input field amplitudes [6], serve as storage for the produced photons and generate time-separation between exiting photons for photon-number-resolved detection [7].

In this paper we present a novel concept of self-phasematched optical frequency conversion in dispersive dielectric microcavities. First a time-dependent model is developed for incorporating the dispersion into the structure of the spatial cavity modes. Then, using this model, actual devices for second harmonic generation (SHG) in integrated semiconductor microcavities are designed and tested experimentally. In a cavity nonlinear interaction, the phase-matching requirement translates into the optimization of a nonlinear mode overlap allowing the design of optimal miniature devices for efficient cavity nonlinear optics.

For the second order $\chi^{(2)}$ nonlinear interaction of SHG the wave equation in M.K.S. units is [8]:

$$\nabla^2 \widetilde{E}_2(\mathbf{r},t) = \mu\varepsilon \frac{\partial^2 \widetilde{E}_2(\mathbf{r},t)}{\partial t^2} + \frac{\chi^{(2)}}{2}\frac{\partial^2 \left(\widetilde{E}_1(\mathbf{r},t)\right)^2}{\partial t^2} \quad (1)$$

where $\widetilde{E}_1(\mathbf{r},t)$ and $\widetilde{E}_2(\mathbf{r},t)$ are the pump and the SH fields respectively.

In contrast to the usual approach of traveling-wave phasematching [1-5,9-14], our model is based on the interaction between spatial cavity eigenmodes in a slowly time-varying envelope approximation. Therefore the pump and the SH fields written as:

$$\widetilde{E}_1(\mathbf{r},t) = E_1(t) e^{i\omega_1 t} F_1(\mathbf{r})$$
$$\widetilde{E}_2(\mathbf{r},t) = E_2(t) e^{i\omega_2 t} F_2(\mathbf{r}) \quad (2)$$

where $F_i(\mathbf{r})$ are the orthonormal set of spatial cavity modes, assumed to be unaffected by the weak nonlinear interaction and $E_i(t)$ are the slowly time-varying envelopes relative to the oscillation period T = 2π/ω, for which the approximation $\frac{\partial E(t)}{\partial t} T \ll E(t)$ can be used.

Assuming a strong undepleted pump $E_1(t)$, the wave equation is

$$E_2(t) e^{i\omega_2 t}\nabla^2 F_2(\mathbf{r}) = \mu\varepsilon_2\left[2i\omega_2 \frac{\partial E_2(t)}{\partial t} - \omega_2^2 E_2(t)\right] e^{i\omega_2 t} F_2(\mathbf{r}) - \mu\frac{\chi^{(2)}}{2}\omega_2^2 E_1^2(t) e^{i\omega_2 t} F_1^2(\mathbf{r}) \quad (3)$$

To calculate the time dependence of the SH mode envelope, the equation is multiplied by $F_2^*(\mathbf{r})$ and integrated yielding

$$E_2(t)\int d\mathbf{r} F_2^*(\mathbf{r})\nabla^2 F_2(\mathbf{r}) = \mu\varepsilon_2 2i\omega_2 \frac{\partial E_2(t)}{\partial t} - \mu\varepsilon_2\omega_2^2 E_2(t) - \gamma\mu\frac{\chi^{(2)}}{2}\omega_2^2 E_1^2(t) \quad (4)$$

where $\gamma = \int d\mathbf{r} F_2^*(\mathbf{r}) F_1^2(\mathbf{r})$ is the nonlinear mode overlap resulting in

$$\frac{\partial E_2(t)}{\partial t} = \frac{1}{2i}\left[\frac{\int d\mathbf{r} F_2^*(\mathbf{r})\nabla^2 F_2(\mathbf{r})}{\mu\varepsilon_2\omega_2} + \omega_2\right] E_2(t) + \frac{i\gamma\chi^{(2)}\omega_2}{4\varepsilon_2} E_1^2(t) \quad (5)$$

The first term on the right-hand side is the cavity loss and the phase dependence designated by $-\frac{\alpha}{2}E_2(t)$ and the equation becomes:

$$\frac{\partial E_2(t)}{\partial t} = \frac{i\gamma\chi^{(2)}\omega_2}{4\varepsilon_2}E_1^2(t) - \frac{\alpha}{2}E_2(t) \qquad (6)$$

Applying zero initial condition for the SH field, the $E_2(t)$ envelope time dependence (for a constant pump) is:

$$E_2(t) = \frac{i\gamma\chi^{(2)}\omega_2}{2\varepsilon_2\alpha}E_1^2\left(1 - e^{-\frac{\alpha}{2}t}\right) \qquad (7)$$

and at steady state, the pump and the SH powers are related by

$$P_2 = \left|\frac{\gamma\chi^{(2)}\omega_2}{2\varepsilon_2\alpha}\right|^2 P_1^2 \qquad (8)$$

Hence in a properly designed cavity, with a large mode overlap γ and small cavity loss term α, the nonlinear process efficiency can be enhanced significantly, regardless of the device dimensions, which can be as small as a few optical wavelengths.

For a doubly-resonant microcavity formed by Bragg mirrors, the optimization of the cavity mode overlap yields different mode sizes for different interacting frequencies. In a one-dimensional case from the traveling-wave point of view, this optimization results in different round-trip times for different cavity modes, and thus it is equivalent to a dispersive reflector design [5] where different phase is accumulated by each frequency in the dispersive mirrors.

We designed an integrated double-resonance microcavity according to the model described above, applying two simultaneous one-dimensional photonic crystal cavities for the pump and for the SH fields. The two periodic structures were applied one on the left side and the other on the right side of a 0.4μm high AlGaInP ridge waveguide (Fig 1-a) with 4 GaInP quantum well (QW) layers. The structure was engineered to have resonances at the 1512nm pump and 756nm SH wavelengths using three-dimensional finite difference time domain (FDTD) simulations taking into consideration the effect of material dispersion. Spatial cavity mode field distributions $F_1(r)$ and $F_2(r)$ were also calculated by the FDTD simulations (Fig 1-b,c).

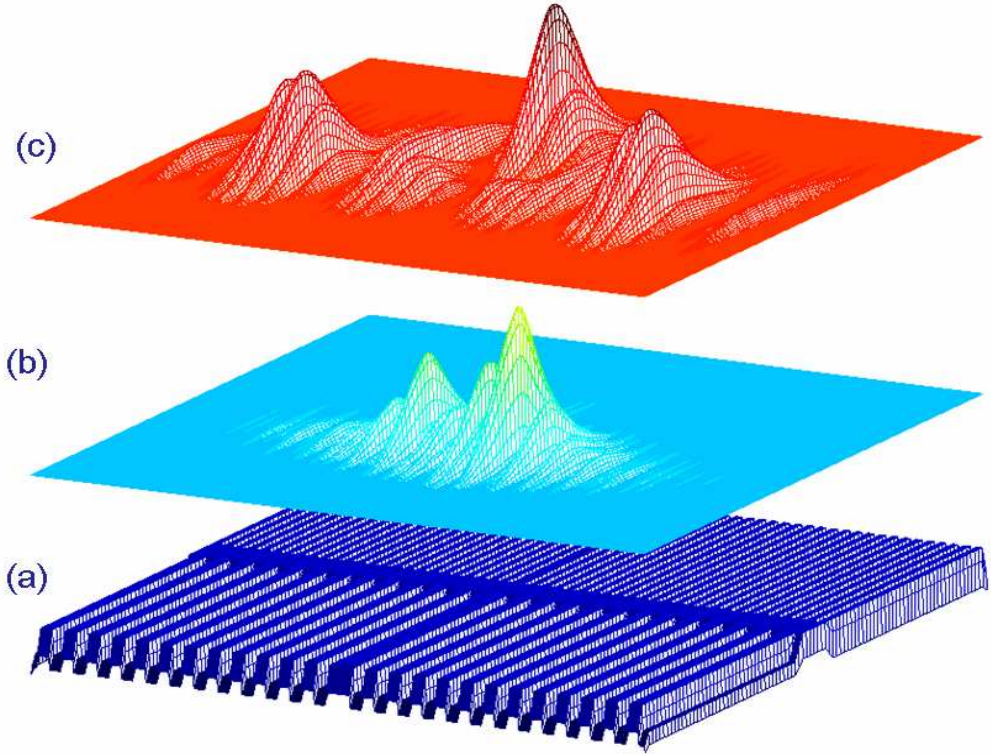

**Fig. 1. (a) – A schematic drawing of the integrated double-resonance microcavity**
**(b) – Calculated (FDTD) SH cavity mode 2D intensity distribution in the QW area [A.U.]**
**(c) – Calculated (FDTD) pump cavity mode 2D intensity distribution in the QW area [A.U.]**

The actual cavities were fabricated on samples consisting of 4 periods of compressively strained 50Å $Ga_{0.45}In_{0.55}P$ QWs separated by 55Å $(Al_{0.5}Ga_{0.5})_{0.51}In_{0.49}P$ barriers. The lateral light confinement was achieved by a 4μm-wide ridge waveguide, realized by etching techniques. The integrated 10μm-long double resonance microcavity was implemented by milling the periodic structure using a FEI Strata 400 Focused Ion beam (FIB) system (Fig 2). The system was operated at an acceleration voltage of 30 kV with a beam dwell time of 1μsec, a 50% overlap and a current of 11pA, corresponding to a beam diameter of 10nm.

The structure was optically pumped by 130fs pulses at telecommunication wavelengths (1500-1530nm) with 80.8MHz repetition-rate using an optical parametric oscillator (OPO), which was pumped by a mode-locked Ti:Sapphire femto-second laser at 810nm. To avoid the adverse effect of two-photon based photoluminescence, special care was taken to perform the measurements far from the resonant two-photon absorption wavelength of $Ga_{0.45}In_{0.55}P$

~1340nm. The OPO output was facet-coupled into the waveguide with about 2μm$^2$ mode area by a polarization maintaining lensed fiber, after filtering out the residual Ti:Sapphire 810nm pump by a thick GaAs wafer. The average pump power used was ~100mW. The generated SH was fiber-coupled into an ANDO spectrum analyzer with optical output, acting as a monochromator, and the photon detection was performed at room temperature by a Perkin-Elmer photon counting module - SPCM 14 which has around 70% efficiency at the SH wavelengths (750-765nm). The measured SHG efficiency - Pout/Pin$^2$ shows a strong maximum near the cavity resonance at ~1512nm (Fig. 3 a) stemming from both the pump power enhancement within the cavity near the resonance and the wavelength detuning effect on mode overlap.

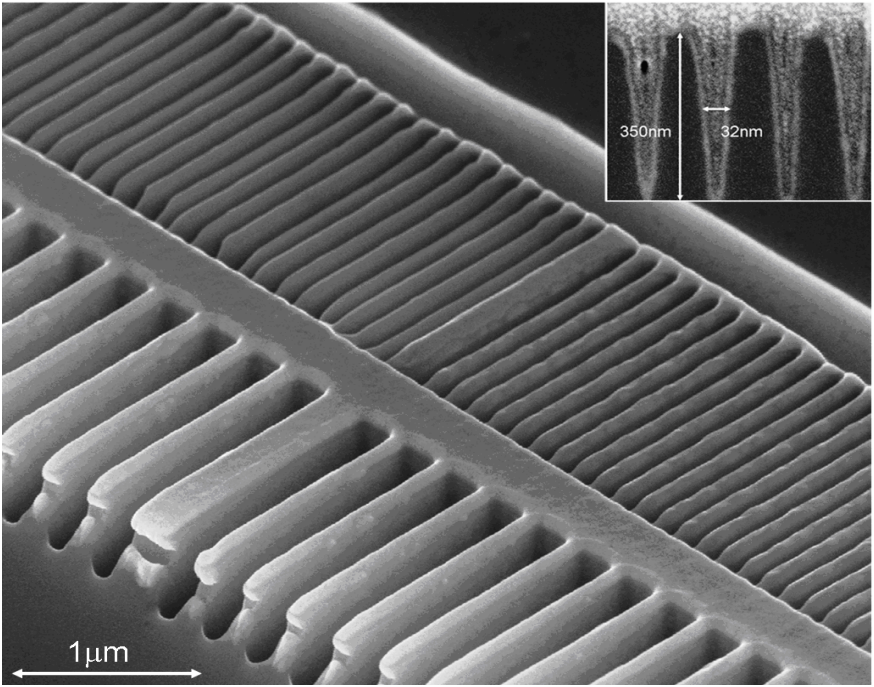

**Fig. 2. Scanning electron microscopy (SEM) image of the fabricated integrated double-resonance microcavity. The inset is a SEM cross-section image of the shortest period grating**.

The effect of wavelength detuning on the nonlinear mode overlap was analytically estimated for a one-dimensional photonic crystal cavity using coupled mode theory for the field in distributed Bragg reflector gratings. The grating complex mode propagation constant is [8]

$$\beta' = \beta_0 \pm i\sqrt{\kappa^2 - (\Delta\beta)^2} \qquad (9)$$

where the detuning $\Delta\beta = \beta(\omega) - \beta_0 = \beta(\omega) - \frac{\pi}{\Lambda}$, $\Lambda$ is the grating period and $\kappa$ is the grating coupling constant.

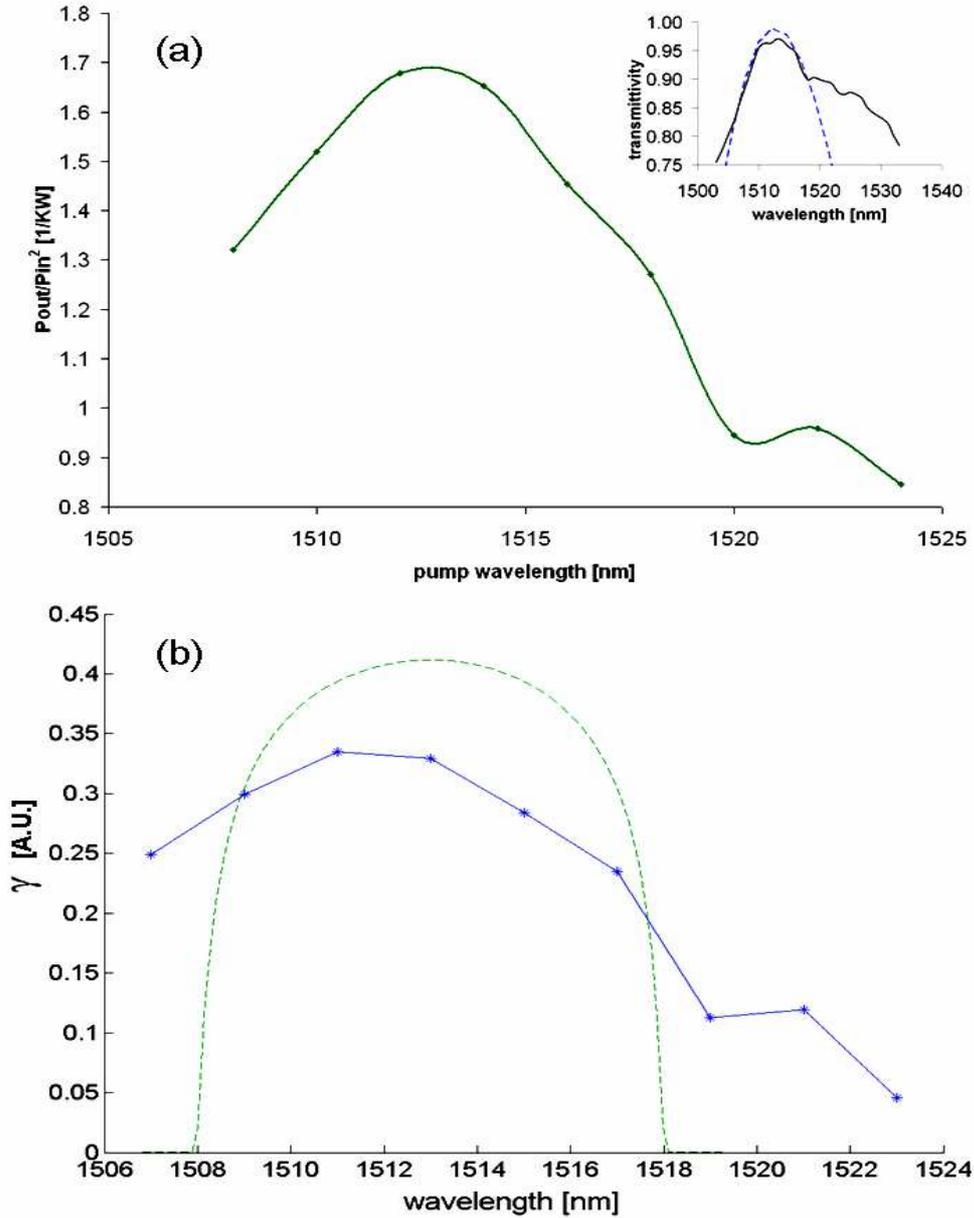

Fig. 3. (a) – Measured SH generation efficiency Pout/Pin$^2$ wavelength dependence. The inset is the microcavity transmission spectrum around the 1512nm resonance. The solid line is the measured and the dashed line is the calculated (FDTD) spectrum.
(b) – Nonlinear mode overlap γ wavelength dependence. The solid line is the measured and the dashed line is the calculated spectrum.

For a small wavelength detuning, the real part of β' in the grating region is determined only by the Bragg period and does not depend on Δβ (Eq. 9). Detuned operation results in slow spatial field decay within the grating region, and thus significant parts of the cavity modes exhibit oscillatory behavior (inside the gratings region). Field oscillation period in the grating region is determined by the real part of β'. Gratings for the pump and the SH fields were designed considering the dispersion, and the resulting spatially oscillating parts of the normalized modes have a small nonlinear overlap. This overlap becomes smaller for a weaker field decay in the grating regions with increasing detuning. The nonlinear mode overlap γ, therefore, has a maximum for $\Delta\beta = 0$ and vanishes for $|\Delta\beta| > |\kappa|$ (Fig3-b). The measured nonlinear mode overlap was extracted using Eq.8 and it clearly resembles the calculated γ dependence on the wavelength detuning.

In conclusion, we have modeled, designed and fabricated integrated 4μm-wide 10μm-long double-resonance semiconductor microcavities for self-phasematched SH generation. The design was based on a theoretical model we have developed for optical frequency conversion in dispersive dielectric microcavities, which includes the dispersion in the structure of the spatial cavity modes. Our experimental results show a strong maximum in the measured SHG efficiency near the cavity resonance due to the increased input power and the wavelength detuning effect on the nonlinear mode overlap, matching our theoretical predictions. The demonstrated integrated microcavity based frequency conversion device can assist in incorporating miniature monolithic semiconductor nonlinear-optic devices into the fields of quantum optics, integrated photonics and optical communications.